\documentclass[a4paper, 10pt, conference]{ieeeconf}      

\usepackage{graphicx}
\graphicspath{{Images/}} 
\usepackage{eso-pic} 
\usepackage{subfig} 
\usepackage{caption} 
\usepackage{transparent}
\usepackage{soul}
\usepackage{amsmath}
\usepackage{framed}

\usepackage{bm}
\usepackage[overload]{empheq}  

\usepackage{tabularx}
\usepackage{longtable} 
\usepackage{colortbl}
\usepackage{booktabs}
\usepackage{multirow}
\usepackage{tablefootnote}

\usepackage{algorithm}
\usepackage{algorithmic}

\usepackage{comment} 
\usepackage{fancyhdr} 
\usepackage{lipsum} 
\usepackage{tcolorbox} 
\usepackage{stfloats} 

\usepackage[colorlinks=true,linkcolor=black,anchorcolor=black,citecolor=black,filecolor=black,menucolor=black,runcolor=black,urlcolor=black]{hyperref} 
\hypersetup{colorlinks=true, allcolors=blue}
\usepackage{cleveref}
\usepackage{soul}

\IEEEoverridecommandlockouts                              
\title{ \bf{Inclusive AI for Group Interactions: Predicting Gaze-Direction Behaviors in People with Intellectual and Developmental Disabilities}}

\begin{document}

\author{
Giulia Huang\\
Politecnico di Milano, Italy\\
giulia.huang@mail.polimi.it
\and
Maristella Matera\\
Politecnico di Milano, Italy\\
maristella.matera@polimi.it
\and
Micol Spitale\\
Politecnico di Milano, Italy\\
micol.spitale@polimi.it
}

\maketitle

\begin{abstract}
Artificial agents that support human group interactions hold great promise, especially in sensitive contexts such as well-being promotion and therapeutic interventions. However, current systems struggle to mediate group interactions involving people who are not neurotypical. This limitation arises because most AI detection models (e.g., for turn-taking) are trained on data from neurotypical populations.
This work takes a step toward \textit{inclusive AI} by addressing the challenge of eye contact detection, a core component of non-verbal communication, with and for people with Intellectual and Developmental Disabilities. 
First, we introduce a new dataset, Multi-party Interaction with Intellectual and Developmental Disabilities (MIDD), capturing atypical gaze and engagement patterns.
Second, we present the results of a comparative analysis with neurotypical datasets, highlighting differences in class imbalance, speaking activity, gaze distribution, and interaction dynamics.
Then, we evaluate classifiers ranging from SVMs to FSFNet, showing that fine-tuning on MIDD improves performance, though notable limitations remain. 
Finally, we present the insights gathered through a focus group with six therapists to interpret our quantitative findings and understand the practical implications of atypical gaze and engagement patterns.
Based on these results, we discuss data-driven strategies and emphasize the importance of feature choice for building more inclusive human-centered tools.
\end{abstract}

\section{Introduction}


Artificial agents that support human group interactions hold great promise, particularly in sensitive contexts such as well-being promotion \cite{spitale2025vita, li2023systematic} and therapeutic interventions \cite{catania2023conversational}. By helping to structure conversations \cite{do2022should}, encourage participation \cite{gillet2022learning}, or mediate turn-taking \cite{jegou2018computational}, these systems can enhance social engagement in group settings. However, most current systems remain narrowly designed for neurotypical populations \cite{whittaker2019disability}. This limitation stems from the training data: nearly all existing AI detection models that detect speaking turns, gaze direction, or attention, are built on corpora reflecting the behaviors of neurotypical participants, which hinders their ability to generalize to heterogeneous and inclusive group settings.

Eye contact, in particular, plays a central role in multi-party social interaction \cite{muller2018robust}, yet the vast majority of AI systems that detect it are trained on datasets reflecting standard social dynamics among neurotypical participants, leading to systematic failures in inclusive contexts. This issue is particularly evident in conversations involving individuals with Intellectual and developmental disabilities (IDDs), lifelong conditions affecting cognitive and adaptive functioning \cite{regier2013dsm}.
Individuals with IDD often exhibit atypical interaction patterns, including reduced or indirect gaze, alternative non-verbal cues (e.g., gestures or vocalizations), and turn-taking rhythms that diverge from neurotypical norms. Moreover, their interactions are frequently mediated by external scaffolds such as caregiver prompts or augmentative communication devices \cite{sigafoos2016augmentative}, which are absent from conventional training data. As a result, AI models for gaze and turn-taking not only lose accuracy but risk marginalizing interaction styles that deviate from the \textit{average} user.

Positioned within the field of human-centered and inclusive AI \cite{caforio2021design}, this paper investigates group dialogues involving participants with IDD. We introduce the Multi-party Interaction with Intellectual and Developmental Disabilities (MIDD) dataset, a carefully annotated corpus that mirrors existing benchmarks such as the MultiMediate dataset \cite{mueller2018lowrapport}, while also capturing the rich behavioral diversity of IDD populations. Building on this resource, we analyze the distributional shifts that distinguish neurodiverse from neurotypical interactions.
To evaluate the impact of these shifts, we benchmark a range of eye-contact detectors, from classical approaches such as Support Vector Classifier (SVC) and eXtreme Gradient Boosting (XGBoost), to the state-of-the-art deep learning model Feature Selection and Fusion Network (FSFNet) \cite{ma2024less}. We assess both zero-shot performance and fine-tuning on MIDD, highlighting improvements as well as persistent limitations.
Finally, we involved six therapists in a focus group to better contextualize our quantitative findings and gather expert insights on interpreting atypical gaze and engagement behaviors.
By documenting behavioral differences (e.g., class imbalance, speaking activity, gaze distribution) and evaluating model robustness, this work contributes empirical insights and a practical foundation for developing inclusive, multimodal eye-contact systems that promote fairness and accessibility in real-world human–computer interaction.

The contributions of this paper are the following:
\begin{itemize}
    \item We introduce a \textbf{novel IDD-focused dataset} (anonymized features are openly available\footnote{https://github.com/giuliahuang/EyeContactModels/tree/main/Dataset}) designed for multi-party group conversations, mirroring the MultiMediate benchmark while capturing unique neurodiverse interaction patterns.
    \item We conduct a rigorous \textbf{comparison analysis} of MIDD with existing datasets (of neurotypical populations), \textbf{highlighting differences in behaviours}, namely gaze behavior, speaking activity, interaction dynamics, and recording quality.
    \item We \textbf{evaluate} performances of both classical (SVC, XGBoost) and state-of-the-art (FSFNet) \textbf{models}, \textbf{showing} their \textbf{limitations in generalising to IDD populations} and the improvements gained via fine-tuning.
    \item We \textbf{incorporate expert insights} from six therapists via a focus group, \textbf{contextualizing quantitative results} and informing the design of inclusive, human-centered interaction systems.
\end{itemize}

This paper provides insights for \textit{inclusive AI} by demonstrating the need for multimodal features (e.g., gaze, head pose) and domain adaptation strategies to accurately model turn-taking and eye-contact in diverse group interactions.

\section{Related Works}

\subsection{Individuals with IDD in group interactions}
Prior research shows that individuals with intellectual and developmental disabilities (IDD) often struggle to claim and maintain turns in multi-party conversations \cite{bigby2014conceptualizing}. Cognitive–linguistic constraints, atypical gaze patterns, and reliance on augmentative communication can reduce verbal initiations and eye contact, while caregivers may unintentionally dominate the dialogue \cite{Antaki2012idadults, Wallin1797526}. Consequently, adults with IDD are at increased risk of marginalization or passive participation in both informal and structured group settings. These difficulties are observed across diverse IDD profiles, including syndromic IDD (e.g., Down syndrome, autism spectrum disorder), neurological impairments, and co-occurring mental health conditions, which can further disrupt conversational coordination. Supporting equitable turn-taking for individuals with IDD is therefore a crucial issue for promoting inclusion and social participation.

These challenges highlight the potential of AI-driven artificial agents to support and moderate group interactions more efficiently. By leveraging real-time detection of gaze, attention, and turn-taking patterns, AI models can guide conversation flow, provide adaptive prompts, promote more balanced participation in multi-party settings.

\subsection{Multi-party turn-taking detection}
Past research has tackled turn-taking in conversations using diverse modalities. Early models focused on textual and syntactic features, e.g., TurnGPT and CNN-FSA hybrids, while later work incorporated prosodic and audio cues (e.g., pitch, speed, volume), yet lacked the ability to infer addressees, limiting their utility in multi-party settings. Visual cues, particularly eye-gaze, have emerged as powerful signals for coordinating turn transitions and managing conversational flow. 
Recent multimodal approaches combine audio, visual, and gestural information, demonstrating improved performance in predicting turn-taking behavior, as exemplified by benchmarks such as the MultiMediate Challenge \cite{mueller2018lowrapport}. Müller et al. \cite{mueller2018eyecontact} further showed that eye-contact detectors can be trained using weak supervision from speaker behavior, enabling unobtrusive gaze modeling. To push state-of-the-art further, FSFNet by Ma et al. \cite{ma2024less} combines CNNs with Transformers and adaptive feature selection to robustly detect eye contact under challenging visual conditions.

Despite these advances, all benchmarks so far are built on neurotypical populations. \textit{No existing work has studied eye-contact behavior in people with Intellectual and Developmental Disabilities} (IDD), despite clear differences in gaze engagement, communication style, and social coordination. This lack of inclusive data and modeling motivates the need for dedicated IDD-focused datasets and robust adaptation strategies for current vision models.

\section{MIDD Dataset}
\label{sec:dataset_collection}
\begin{figure}[htbp]
    \centering
    \includegraphics[width=\columnwidth]{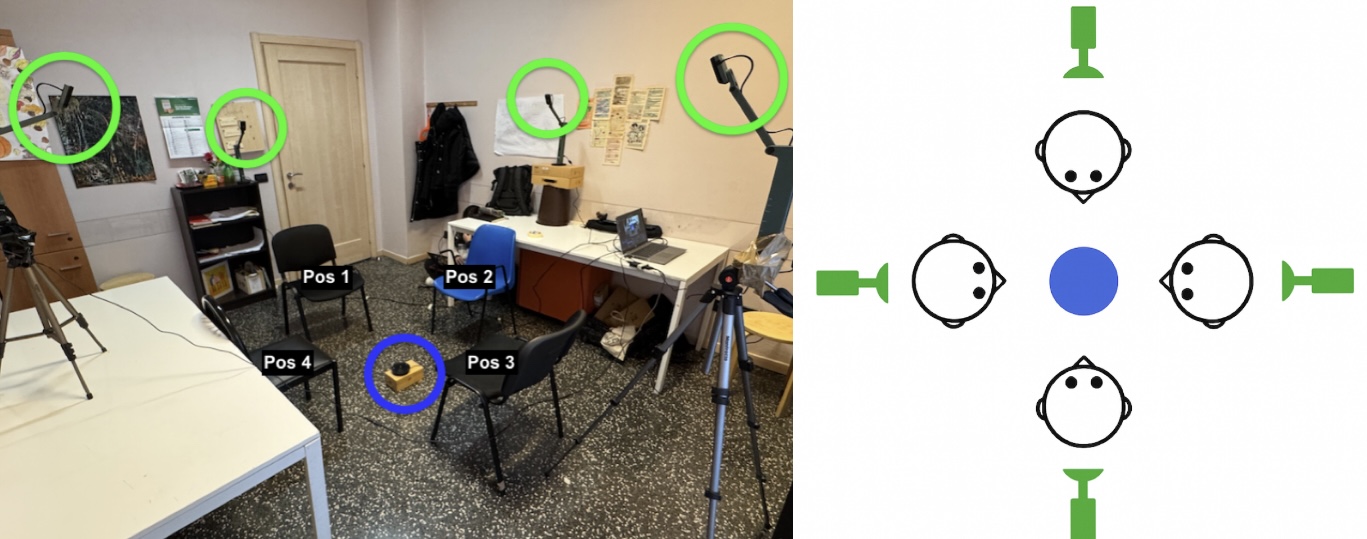}
    \caption{Illustration of the recording setup for the MIDD dataset with position labels. Cameras are indicated with green, microphone with blue.}
    \label{fig:recording_setting}
\end{figure}

\noindent
To address the lack of corpora involving groups of adults with IDD, we collected the Multi-party Interaction with Intellectual and Developmental Disabilities (MIDD) dataset, a set of multi-party conversations designed to mirror the MultiMediate benchmark \cite{mueller2018lowrapport} while capturing the distinctive communicative behaviours of this under-represented population. 
Thirteen adults (9 female, 4 male, between 23 and 51 years) with heterogeneous IDD profiles, including moderate/severe intellectual disability, Down syndrome, autism, and comorbid neuropsychiatric conditions, took part in six group interactions (five four-person groups and one five-person group) recorded across two sessions. Each group was formed by the educators of the daycare centre \textit{[omitted]}, who were responsible for organizing participants in these groups to include individuals with varying abilities and cognitive levels. Each session lasted approximately 20 minutes and was filmed in a quiet, semi-circular seating arrangement using four synchronised webcams with a single central microphone, replicating the original benchmark’s room layout \footnote{\url{https://multimediate-challenge.org/datasets/Dataset_MPII/}} for direct comparability (as shown in Figure \ref{fig:recording_setting}). A trained educator of the daycare centre guided semi-structured discussions on personally meaningful topics (e.g., Christmas, friends) to stimulate natural gaze, turn-taking, and engagement patterns. 
All participants’ legal guardians provided informed consent for their involvement in the study, and participants were free to withdraw at any time. The data collection was approved by the Ethics Committee of Politecnico di Milano University.


\textbf{Features set.} The resulting MIDD corpus comprises 24 synchronised videos ($\approx 476\ \mathrm{min}$) of six multi-party conversations. Each recording captures a single focal participant but was manually cropped to minimise background faces, then segmented into non-overlapping 10-second chunks that inherit the MultiMediate format.  From every segment we extracted the final frame and used OpenFace 2.0 \cite{openface} to obtain aligned $112\times112$ facial images alongside gaze and head-pose vectors; missing detections were replaced by black frames or zero vectors. 

\textbf{Annotation.} Segments were subsequently labelled for (i) relative eye-contact direction (none, left, front, right) with respect to the participant and (ii) active-speaker identity. Initial speaker diarization masks from \texttt{pyannote.audio}\footnote{https://github.com/pyannote/pyannote-audio} were refined in ELAN\footnote{https://archive.mpi.nl/tla/elan} to ensure frame-accurate alignment and supplemented with binary and one-hot speaker indicators.
To evaluate annotation reliability, two independent annotators were involved: one annotated the full dataset, while the second annotated a randomly selected 20\% subset. The resulting inter-annotator agreement was 82\%, indicating substantial consistency across labels. An analysis of disagreement ratios (defined as disagreements over total labels per class) revealed that segments without eye contact were slightly more ambiguous (57/305, 0.187) than those with eye contact (27/165, 0.164), suggesting that the absence of clear gaze cues poses greater annotation difficulty.
Annotating the MIDD corpus presented practical challenges. Ambiguous gaze toward non-participants, suboptimal lighting, and participants’ occluding behaviours complicated eye-contact labeling. Additionally, the use of a single central microphone resulted in noisy, overlapping speech that required extensive manual correction after applying speech diarization.

\textbf{Samples.} The final dataset (Table \ref{tab:class-distribution}) is composed by 2\,962 samples whose eye-contact labels are highly imbalanced (66.6\% “no eye contact”). One full recording session as an independent 610-sample test set, while the remaining data were used in a five-fold cross-validation setup to produce training and validation splits (1\,882/470 samples) for model development.

Overall, the MIDD dataset offers rich, dual-labelled evidence of gaze, pose, and speech behaviours in ecologically valid group settings, thereby enabling rigorous evaluation and adaptation of eye-contact detection models to neurodiverse interactions. The anonymized features set is publicly available on GitHub\footnote{https://github.com/giuliahuang/EyeContactModels}.

\begin{table}[h]
\centering
\caption{Class distribution of eye contact labels in the MIDD Dataset}
\label{tab:class-distribution}
\begin{tabular}{cccc}
\toprule
\textbf{Class Label} & \textbf{Direction}            & \textbf{Count} & \textbf{\%} \\
\midrule
0 & None     & 1,972 & 66.6\% \\
1 & Right    & 342   & 11.5\% \\
2 & Front    & 334   & 11.3\% \\
3 & Left     & 314   & 10.6\% \\
\midrule
\textbf{Total} &                                & \textbf{2,962} & \textbf{100\%} \\
\bottomrule
\end{tabular}
\end{table}

\section{Analysis of Neurotypical vs. IDD Datasets}

\noindent
To understand how interaction patterns and recording conditions differ between neurotypical participants and those with Intellectual and Developmental Disabilities (IDD), we directly compare the MIDD dataset to the established MultiMediate corpus. 
The MIDD dataset was designed to replicate the MultiMediate setup as closely as possible, matching group size, recording environment, and 10-second video chunk length, and uses the same labeling convention (eye contact and speaking labels refer to the last frame of each chunk).  
We evaluate three facets of group conversation: (1) \emph{engagement metrics}, counts of eye contact, per-participant speaking durations, and listener focus during active turns; (2) \emph{behavioral dynamics}, turn-taking balance and initiation of non-verbal cues; and (3) \emph{image quality}, brightness and contrast variation under uncontrolled lighting. 
This controlled comparison is crucial for (a) revealing the distinctive communicative behaviors of adults with IDD in multi-party settings \cite{antaki2017profound} \cite{Tepencelik2024autism}, (b) diagnosing why models trained on high-quality, neurotypical data struggle to generalize to atypical interaction styles and variable visuals, and (c) motivating the design of multimodal, robust solutions using diverse modalities \cite{mazzola2023icub} that accommodate both unconventional social signals and lower-quality imagery.

\begin{figure}[h]
    \centering
    \includegraphics[width=\columnwidth]{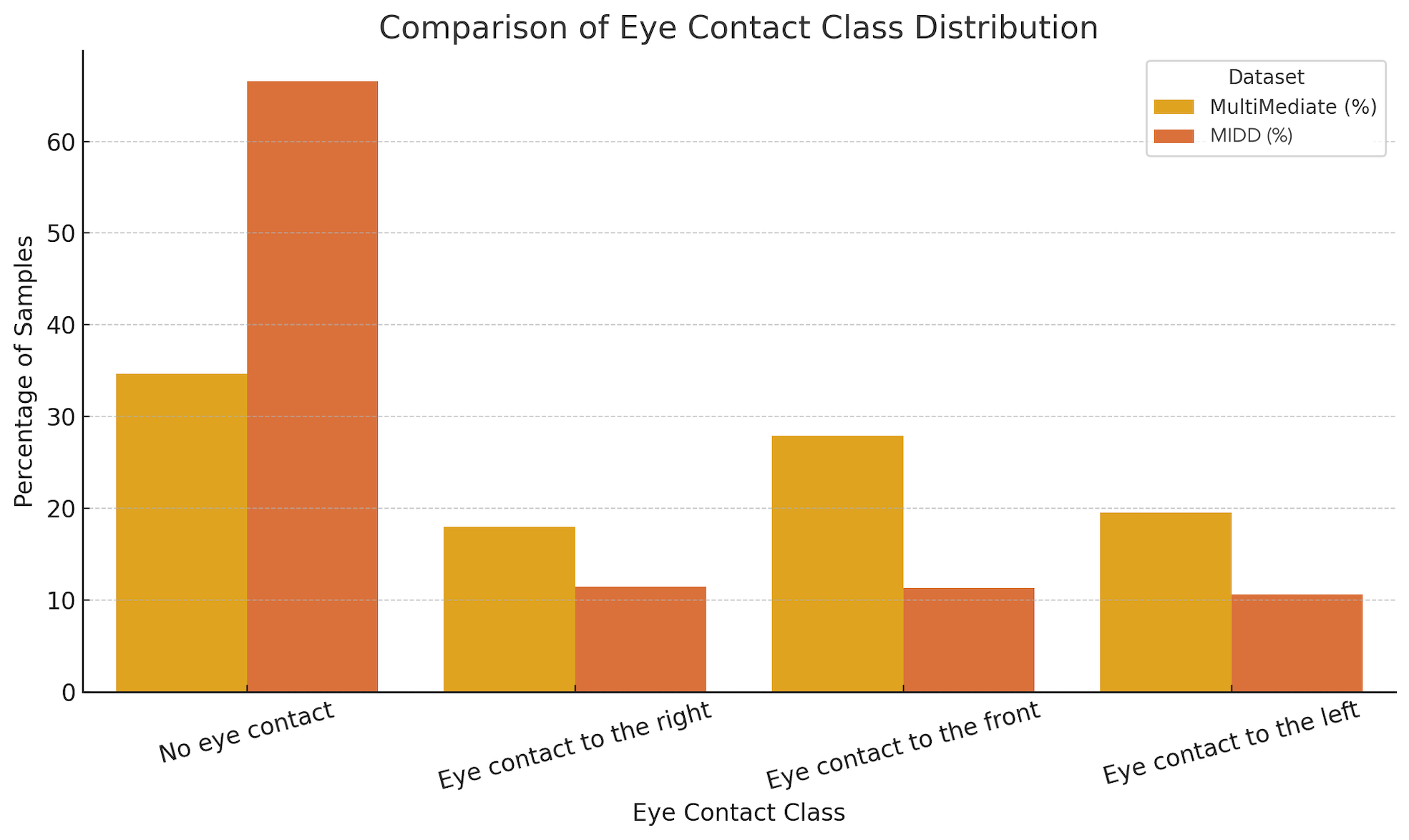}
    \caption{Eye contact class distribution comparison: MIDD vs. MultiMediate}
    \label{fig:class_distribution_comparison}
\end{figure}

Although both corpora share identical group size, 10s chunking, and annotation schemes, MIDD is dominated by ‘‘no eye contact’’ labels, as shown in Figure \ref{fig:class_distribution_comparison} (66.6\% vs.\ 34.7\%), and overall gaze engagement is almost halved (33\% vs.\ 65\%). A chi-square test of independence (\(\chi^{2}(1)=429.4, p<.001\)) confirms that the disparity in eye-contact frequencies is far too large to arise by chance, with one degree of freedom reflecting the two categories (‘‘eye-contact" vs. ‘‘no eye-contact").

Speaking behaviour diverges similarly. Active-speaker events, where at least one participant is talking, occur in only 62\% of MIDD samples compared with 92\% in MultiMediate. Again, a chi-square test (\(\chi^{2}(1)\approx 500, p<.001\)) indicates an extremely significant difference. Moreover, the distribution of relative speaking time across participants is noticeably more unequal in MIDD (Gini =.52) than in MultiMediate (Gini = .34), quantifying a higher concentration of speaking among a few individuals. Note that we used the Gini coefficient to measure the inequality among the behavioral cues of the neurotypical frequency distribution \cite{dorfman1979formula}.

When focusing on speech segments (Table \ref{tab:speaker_eye_contact}), IDD speakers establish eye contact far less often (35\% vs.\ 64\%, \(p<.001\) by chi-square), yet listener attention to the current speaker shows in mild, non-significant increase in MIDD  (53\% vs.\ 48\%, \(p=.087\)), suggesting a trend but one that does not meet the conventional \(\alpha = .05\) threshold. 

\begin{table}[h]
\centering
\footnotesize
\caption{Proportion of speaker eye contact during speaking} 
\label{tab:speaker_eye_contact}
\begin{tabular}{lccc}
\toprule
\textbf{Dataset} & \textbf{Eye contacts} & \textbf{Speaking Events} & \textbf{Rate} \\
\midrule
MultiMediate & 1,043 & 1,626 & 64.2\% \\
MIDD         & 128   & 365   & 35.1\% \\
\bottomrule
\end{tabular}
\end{table}

To assess listener engagement during conversations, we analyzed how frequently at least one listener made eye contact with the active speaker during speaking turns. Contrary to speaker behavior, which showed less eye contact in the MIDD dataset, listeners in MIDD appeared slightly more engaged: 53.15\% of speaking events involved at least one listener looking at the speaker, compared to 48.03\% in the MultiMediate dataset. However, a Chi-squared test ($p = .087$) indicates that this difference is not statistically significant. These results suggest a mild, non-significant trend of increased listener attentiveness in MIDD, possibly pointing to different social dynamics or reliance on visual cues for coordination.

A quantitative comparison of frame brightness and contrast confirms that the MIDD corpus poses a significantly greater visual challenge: its brightness and contrast distributions are broader and more skewerd, with many under-exposed, low-variance frames. Mann-Whitney U tests on these image quality metrics both yield \(p<.001\), indicating that every pairwise brightness and contrast comparison differs at a high level of confidence.

Within the MIDD corpus, we also investigated whether speaking-time share correlates with the coded severity of intellectual disability.  After normalizing each participant’s total speaking duration and encoding severity on a three-level ordinal scale (Moderate = 2, Moderate–Severe = 3, Severe = 4), Pearson's correlation \(r=-0.08\) and Spearman’s rank correlation \(\rho=-0.12\,(p=.74)\) both fail to reach significance, revealing only a weak, non-reliable tendency for more severely impaired individuals to speak less. Descriptively, the Severe group contributes the least (2-9\% of total speaking time), whereas two highly engaged Moderate-Severe participants elevate that category's mean, illustrating how individual variability and comorbidities can outweigh nominal severity labels.

\begin{framed}
\noindent
\textbf{Key findings:} Overall, these results document pronounced reductions in both gaze and verbal engagement, increased speaking-time inequality, altered listener–speaker dynamics, and degraded image quality in the MIDD corpus. Any robust eye-contact detection model must therefore address severe class imbalance, heterogeneous participant behaviours, and variable visual conditions, most likely via class-balanced training strategies and multimodal feature integration.
\end{framed}


\section{Eye Contact Modeling}
\label{sec:experiments}%


\subsection{Datasets}

\textbf{MIDD Dataset.} The MIDD Dataset, introduced in Section \ref{sec:dataset_collection}, comprises 24 synchronized videos of approximately 20 minutes each, capturing six conversations involving thirteen Italian adults with IDD. Each conversation group consisted of 4-5 participants engaged in predefined, topic-free discussions. To align with the structure of the MultiMediate dataset, each video was segmented into non-overlapping 10-second chunks. From every chunk, the last frame was extracted, yielding a total of 2,962 samples. These were then divided into 1,882 training, 470 validation and 610 test samples. 

\textbf{MultiMediate Challenge.} The MultiMediate Challenge dataset includes recordings of 78 university students (43 female), aged between 18 and 38 years. These participants were organized into 12 four-person groups and 10 three-person groups. Each conversation was recorded by 8 frame-synchronised video cameras and 4 microphones and each group was asked to select and discuss a controversial topic from a predefined list, resulting in spontaneous, unscripted social interactions. The dataset consists of 10-second video chunks, already partitioned into training and validation sets. These samples were drawn from 22 distinct recording sessions, with 16 allocated to the training set and 6 to the validation set. In total, the training set comprises 4,505 samples, while the validation set includes 1,673 samples.

Samples from both datasets are re‐processed with OpenFace to yield aligned $112\times112$ face crops, twelve numerical gaze–pose features, and six manually annotated speaker‐related columns.  The gaze features (\texttt{gaze\_0\_x}, \texttt{gaze\_0\_y}, \texttt{gaze\_0\_z}, \texttt{gaze\_1\_x}, \texttt{gaze\_1\_y}, and \texttt{gaze\_1\_z}) represent normalized direction vectors in world coordinates for the left and right eyes.  The pose features consist of three translation components (\texttt{pose\_Tx}, \texttt{pose\_Ty}, \texttt{pose\_Tz}), denoting 3D head position relative to the camera in millimetres, and three rotation components (\texttt{pose\_Rx}, \texttt{pose\_Ry}, \texttt{pose\_Rz}), denoting head rotation about the X, Y, and Z axes in radians.  The speaker‐related annotations include \texttt{is\_speaker}, a binary flag indicating whether the participant was speaking, and \texttt{speaker\_info}, the speaker’s identity (0–4) encoded as five one‐hot binary columns. Altogether, this yields an 18-dimensional feature vector per frame.

\subsection{Models}

\textbf{SVC.} The Support Vector Classifier (SVC) is a traditional machine learning model trained on handcrafted gaze and head pose features extracted via OpenFace, following Müller et al.'s baseline setup. Specifically, it uses 18-dimensional vectors combining 3D head translations and rotations, gaze direction vectors from both eyes and speaking cues. After removing missing values and standardizing the features, a grid search over $C$ and $\gamma$ was performed using RBF kernels to tune the model. 

\textbf{XGBoost.} XGBoost is a gradient-boosted tree ensemble known for efficient handling of tabular data and strong performance on imbalanced classification tasks. The model was trained on the same 18-dimensional feature vectors used for SVC (gaze directions, head pose estimates from OpenFace and speaking cues) and configured to learn subject-agnostic gaze targets. Prior to training, all features were standardized and missing entries removed. A grid search over hyperparameters (tree depth, learning rate, number of estimators, regularization) was conducted using GPU acceleration. 

\textbf{FSFNet.} FSFNet is a transformer-based architecture tailored for multi-party eye contact detection. It takes as input a $112 \times 112$ RGB face crop and outputs a 4-class probability distribution indicating gaze direction (no contact, right, front, left). The network consists of a truncated ResNet-50 backbone for spatial feature extraction, followed by a transformer encoder (FSFEncoder) that performs attention-based token pruning, and a classification head. A channel attention module enhances important features before classification \cite{ma2024less}. 

\subsection{Experimental Setup}

To systematically evaluate eye-contact detection in multi-party interactions involving adults with intellectual and developmental disabilities (IDD), we adopted a multi-model experimental framework combining both vision-based deep learning and structured-feature classifiers. 
All models were trained on the dataset of 2,962 annotated samples drawn from our proposed MIDD corpus, with the following fixed split: 1,882 training, 470 validation, and 610 test samples. Labels range from 0 to 3 and encode four eye-contact categories: no contact, right gaze, front gaze, and left gaze.

The SVC and XGBoost models were trained on a larger combined dataset (8,058 training samples), created by merging the MIDD training set with samples from the MultiMediate corpus. This hybrid training set aimed to encourage \emph{cross-domain generalization} while preserving task-specific performance on IDD data. A stratified validation set of 470 samples from MIDD was retained to evaluate performance in the target domain.
For FSFNet, we adopted a \emph{transfer learning} approach: pretrained weights from the MultiMediate benchmark were used to initialize the model, which was then fine-tuned on the MIDD data using the original hyperparameter settings (Adam optimizer, learning rate, batch size). To address class imbalance, we experimented with PyTorch’s \texttt{WeightedRandomSampler}, ensuring rare gaze classes were sampled more frequently during training.

In addition to standalone evaluations, we proposed \textbf{ensemble strategies} to combine complementary information across modalities. Two strategies were employed: (1) Weighted ensembling, where final predictions were computed as a convex combination of FSFNet and either SVC or XGBoost outputs, with weights chosen empirically (e.g., 0.6 FSFNet / 0.4 XGBoost); and (2)  Stacking, where predictions from the base models served as features for a logistic regression meta-classifier trained to optimize final output decisions.
An \emph{ablation study} was also conducted for SVC and XGBoost, removing feature groups (gaze, pose, speaker cues) individually to assess their relative contribution.
Performance was evaluated using three complementary metrics: (i) Classification Accuracy, measuring overall correctness; (ii) Macro-averaged F1 Score, to assess per-class performance in the presence of imbalance; Composite Score\footnote{https://github.com/ma-hnu/FSFNet/}, defined a weighted combination of macro-averaged $F_1$ score and overall accuracy \cite{ma2024less}.
All metrics were computed independently on the training, validation, and test sets, enabling a direct comparison of generalization capabilities across model types, training strategies, and fusion schemes.


\subsection{Modeling Results}
We evaluated three eye-contact detectors (FSFNet, SVC and XGBoost) on the MultiMediate benchmark and prepared them for subsequent fine-tuning on the MIDD corpus.  
FSFNet, a ResNet-50–based transformer with sharpness-aware minimisation, reached state-of-the-art performance after 42 of 300 epochs (82.1\% validation accuracy, 89.1\% training accuracy, final loss .33).  The handcrafted-feature baseline, an RBF-kernel Support Vector Classifier (SVC) \cite{mueller2018eyecontact} tuned via grid search, achieved 63\% training and 56.8\% validation accuracy (macro-F1 0.60), while a GPU-accelerated XGBoost model improved these scores to 73\% training and 62\% validation accuracy (macro-F1 0.64), both trained on gaze and pose related features. This establishes a controlled hierarchy of deep and tree-based models whose relative strengths (XGBoost’s class-imbalance robustness, SVC’s limited generalisation) guide subsequent fine-tuning on the MIDD corpus.
Information about the hyperparameters, checkpoint of the models, and the corresponding code can be found on GitHub\footnote{https://github.com/giuliahuang/EyeContactModels}.

\begin{table}[htbp]
\centering
\footnotesize
\caption{Training and Validation Accuracy for FSFNet, SVC, and XGBoost on MultiMediate Dataset}
\label{tab:model_acc}
\begin{tabular}{lcc}
\toprule
\textbf{Model} & \textbf{Train Accuracy} & \textbf{Val Accuracy} \\ 
\midrule
\textbf{FSFNet}    & \textbf{.89} & \textbf{.82} \\
SVC       & .63 & .57 \\
XGBoost   & .73 & .62 \\
\bottomrule
\end{tabular}
\end{table}

\begin{table}[htbp]
\centering
\footnotesize
\caption{Performance of FSFNet, SVC, and XGBoost on the MIDD Dataset}
\label{tab:train_val_midd}
\begin{tabular}{lcccc}
\toprule
\textbf{Model} & \textbf{Train Acc.} & \textbf{Val Acc.} & \textbf{F1 Score} & \textbf{Score} \\
\midrule
FSFNet (base)               & .64 & .82 & .71 & .75 \\
FSFNet (aug)                & .60 & .82 & .74 & .76 \\
SVC                         & .67 & .69 & .44 & .52 \\
XGBoost                     & .43 & .70 & .40 & .50 \\
\bottomrule
\end{tabular}
\end{table}

\begin{table*}[htbp]
\centering
\footnotesize
\caption{Validation performance of SVC and XGBoost across feature subsets}
\label{tab:svc_xgb_ablate}
\begin{tabular}{@{}lcccccc@{}}
\toprule
\multirow{2}{*}{\textbf{Features}} &
\multicolumn{3}{c}{\textbf{SVC}} &
\multicolumn{3}{c}{\textbf{XGBoost}} \\
\cmidrule(lr){2-4} \cmidrule(lr){5-7}
& \textbf{Acc} & \textbf{Macro $F_1$} & \textbf{Score}
& \textbf{Acc} & \textbf{Macro $F_1$} & \textbf{Score} \\
\midrule
\{\textit{gaze\_of}\} & .61 & .27 & .38 & .62 & .25 & .37 \\
\{\textit{gaze\_of}, \textit{is\_speaker}\} & .66 & .25 & .39 & .61 & .25 & .37 \\
\{\textit{gaze\_of}, \textit{is\_speaker}, \textit{pose\_of}\} & .67 & .37 & .47 & .68 & .37 & .47 \\
\{\textit{gaze\_of}, \textit{pose\_of}\} & .69 & .43 & .52 & .68 & .37 & .47 \\
\{\textit{is\_speaker}, \textit{pose\_of}\} & .67 & .31 & .43 & .69 & .40 & .50 \\
\{\textit{pose\_of}\} & .67 & .31 & .43 & \textbf{.70} & \textbf{.40} & \textbf{.50} \\
\{\textit{gaze\_of}, \textit{is\_speaker}, \textit{pose\_of}, \textit{speaker\_info}\} & \textbf{.69} & \textbf{.44} & \textbf{.52} & .67 & .35 & .46 \\
\{\textit{gaze\_of}, \textit{is\_speaker}, \textit{speaker\_info}\} & .66 & .32 & .43 & .61 & .29 & .40 \\
\{\textit{gaze\_of}, \textit{pose\_of}, \textit{speaker\_info}\} & .67 & .43 & .51 & .67 & .35 & .46 \\
\{\textit{pose\_of}, \textit{speaker\_info}\} & .67 & .28 & .41 & .68 & .37 & .47 \\
\{\textit{gaze\_of}, \textit{speaker\_info}\} & .66 & .24 & .38 & .61 & .29 & .40 \\
\{\textit{is\_speaker}, \textit{pose\_of}, \textit{speaker\_info}\} & .68 & .36 & .47 & .68 & .37 & .47 \\
\bottomrule
\end{tabular}
\end{table*}

\noindent
To assess domain adaptation, we fine-tuned three architectures, FSFNet, an RBF-kernel Support Vector Classifier, and a GPU-accelerated XGBoost, on the imbalanced MIDD corpus, summarized on Table \ref{tab:train_val_midd}.  For FSFNet we applied five-fold stratified cross-validation, obtaining a highest accuracy of~$\approx\!.82$ and macro $F_{1}\approx.71$; class-balanced oversampling lifted minority-class $F_{1}$ while maintaining comparable accuracy, but further full fine-tuning degraded overall performance, evidencing mild overfitting.  In contrast, the SVC, even when trained on a concatenation of MultiMediate and MIDD data with multimodal gaze, pose, and speaker features, defaulted to predicting the majority “no-eye-contact’’ class, yielding respectable accuracy ($.69$) yet a macro $F_{1}$ of only~$.44$.  XGBoost showed a different bias: after fine-tuning with speaker cues, it reached $.62$ accuracy and macro $F_{1}=.30$, excelling at class~0 (precision/recall~$.69/.91$) but sacrificing recall on true gaze events. Overall, deep transfer learning plus cautious oversampling offers the best balance between majority precision and minority recall, whereas classical models remain restricted by extreme class imbalance, highlighting the need for cost-sensitive objectives and richer augmentation to achieve equitable eye-contact detection in neurodiverse group interactions.  
 
\noindent
Finally, we ensembled the vision-based FSFNet (kept in its original pre-trained form) with the best classical models. Averaging FSFNet with an SVC trained on gaze + speaker cues (optimal weight 0.7/0.3) raised accuracy to \(\,.67\), macro \(F_{1}\) to \(.54\), and preserved FSFNet’s precision on the majority “no-contact’’ class while noticeably lifting recall for minority gaze directions.  A second ensemble pairing FSFNet with an XGBoost model achieved the strongest results at a 0.6/0.4 weighting, attaining \(.70\) accuracy, macro \(F_{1}=.55\), and a composite score of \(.60\); per-class analysis shows sizeable gains for classes 1 and 2 without sacrificing performance on class 0. A logistic-regression stacker offered similar but slightly lower scores, confirming that simple averaging suffices to harness complementary strengths and deliver the most balanced eye-contact detection to date on neurodiverse interactions. To assess the statistical significance of these improvements, we conducted McNemar’s tests comparing the best ensemble model against the two base models. Results showed significant differences in prediction errors between the ensemble and FSFNet (original) (\(\chi^2 = 18.62\), \(p < .001\)), as well as between the ensemble and XGBoost (\(\chi^2 = 12.82\), \(p < .001\)), confirming the ensemble’s superiority on the test set. The performance metrics for each model are summarized in Table~\ref{tab:models_test_evaluation}.

\begin{table}[htbp]
\centering
\footnotesize
\caption{MIDD Test‐Set Performance for Individual Models and Ensembles}
\label{tab:models_test_evaluation}
\begin{tabular}{@{}llccc@{}}
\toprule
\multicolumn{2}{l}{\textbf{Model}} & \textbf{Acc} & \textbf{Macro \(F_{1}\)} & \textbf{Score} \\ 
\midrule
\multicolumn{2}{l}{\textit{Individual Models}} \\
& FSFNet (original) & .64 & .53 & .57 \\
& SVC & .60 & .29 & .39 \\
& XGBoost & .62 & .30 & .40 \\
\addlinespace
\multicolumn{2}{l}{\textit{FSFNet + SVC}} \\
& Ensemble (F$_{0.7}$/S$_{0.3}$)   & .67 & .54 & .58 \\
& Stacked (F + S)                  & .66 & .54 & .58 \\
\addlinespace
\multicolumn{2}{l}{\textit{FSFNet + XGBoost}} \\
& Ensemble (F$_{0.6}$/X$_{0.4}$)   & \textbf{.70} & \textbf{.55} & \textbf{.60} \\
& Stacked (F + X)                  & .67 & .54 & .59 \\
\bottomrule
\end{tabular}

\medskip
S = SVC trained on \{\texttt{gaze}, \texttt{is\_speaker}, \texttt{speaker\_info}\};   
X = XGBoost trained on the same features;   
F$_{w}$/S$_{1-w}$ and F$_{w}$/X$_{1-w}$ denote weighted‐average ensembles;   
Stacked refers to a logistic‐regression meta‐classifier.  
\end{table}


\begin{framed}
\noindent
\textbf{Model performance summary:} A simple weighted average ensemble of FSFNet’s visual outputs and SVC’s gaze–speaker signals achieves superior performance. It reached 70\% in accuracy and enhanced balance in both \(F_{1}\) (55\%) and composite score (60\%), compared to each standalone model and a stacked meta-classifier, demonstrating the efficacy of averaging to capture diverse turn-taking cues in IDD conversations.
\end{framed}



\subsection{Ablation Study}
\label{section:ablation_study}

To investigate the impact of different input signals on model performance, we conducted an ablation study by training both SVC and XGBoost models on all 12 combinations of gaze, pose, and speaker-activity features. Validation results (Tables~\ref{tab:svc_xgb_ablate}) reveal that configurations including pose (\texttt{pose\_of}) yielded the highest accuracy and $F_1$ scores on the validation set for both models—suggesting that pose carries strong discriminative power within the development split.

However, when these pose-driven models were applied to the unseen MIDD test set, their performance collapsed: they predicted nearly exclusively the dominant class (0: no eye contact), leading to sharp drops in macro $F_1$ (as low as .20 for SVC and .24 for XGBoost). This overfitting behavior is especially problematic in imbalanced multi-class settings, where high accuracy can mask poor minority-class recovery.

In contrast, feature subsets that excluded pose and relied solely on gaze and speaker-activity cues (\{\texttt{gaze\_of}, \texttt{is\_speaker}, \texttt{speaker\_info}\}) generalized more robustly. On the test set, both classifiers trained on this configuration achieved a macro $F_1$ around 0.27–0.30, recovering non-trivial proportions of minority classes while maintaining competitive overall accuracy. Among these, the best-performing models on test data were: (i) the XGBoost classifier trained on \{\texttt{gaze\_of}, \texttt{is\_speaker}, \texttt{speaker\_info}\}, with $\text{Acc}=.62$, macro $F_1=.30$, and Composite Score $=.41$; and (ii) the SVC variant using the same feature set, with $\text{Acc}=.60$, macro $F_1=.27$, and Composite Score $=.38$.

A further contributor to this discrepancy is the relabeling of the MIDD test set, which contains a five-person conversation while the model supports only four target classes. To enable inference, ground-truth labels were remapped from five relative eye-contact directions to four supported classes, introducing semantic ambiguity—e.g., class 1 (rightward gaze) could correspond to two different individuals. As pose-based models rely on spatial alignment between head orientation and speaker identity, this mismatch disrupted their assumptions and led to misclassifications, highlighting their sensitivity to inconsistencies in spatial layout and label semantics. Gaze and speaker cues, while weaker in isolation, yield more balanced predictions across classes, likely due to their stronger invariance to domain shifts.

\subsection{Discussion and Limitations}
Despite solid performance on the benchmark MultiMediate dataset, all our models (FSFNet, SVC, and XGBoost) experienced sharp performance degradation when transferred to the MIDD dataset. We attribute this to five main causes. First, the proposed dataset involves a relatively small number of participants (13 in total). This limited sample size restricts the diversity of interaction patterns represented in the data and constraints the kinds of insights that can be drawn. Second, the fidelity of the input data varies substantially: MIDD frames suffer from poor lighting and varied camera angles, which impair facial landmark and gaze estimation. Moreover, the MIDD dataset has a heavy class imbalance, with over 60\% of frames labeled as “no eye contact,” biasing learning dynamics. Third, the behavioral diversity across MIDD participants, including spanning autism, Down syndrome, and cerebral palsy, introduces variability in how turn-taking is signaled, often through non-standard or non-visual modalities. This severely limits the effectiveness of models trained on neurotypical data. Fourth, feature selection significantly impacted performance: pose-based models showed strong validation performance but collapsed to predicting only the majority class at test time. In contrast, configurations that excluded pose and emphasized direct gaze and speaker-state information yielded more balanced predictions, achieving better macro \(F_1\) scores despite slightly lower accuracy. Last, temporal information is only considered within a fixed window (10 seconds), and the baseline methods are limited to FSFNet and a few additional ML approaches. More advanced temporal models, which could better capture the complexity of interactions in MIDD, were not explored.

Our findings reveal the limitations of direct model transfer from neurotypical to IDD populations and highlight the need for domain-adaptive modeling approaches. Future research should investigate self-supervised pretraining, multimodal feature fusion, and context-aware modeling to capture the richness and variability of communication in IDD settings.

\begin{table*}[h!]
\centering
\footnotesize
\caption{Mapping of therapist insights to quantitative findings}
\begin{tabular}{p{5cm} p{11.5cm}}
\hline
\textbf{Therapist insight} & \textbf{Relation to quantitative findings} \\
\hline
Heterogeneous interaction patterns & High variability in speaking activity and turn-taking; some participants dominate while others remain passive, explaining skewed distributions and class imbalances. \\
\hline
Critical role of scaffolding & Educators guide turn-taking through gaze and posture; aligns with higher engagement or gaze alignment in certain sessions, suggesting models should account for contextual support. \\
\hline
Multimodal communication & Reliance on gestures, facial expressions, symbolic tools, and touch indicates that gaze-only models may miss meaningful communication, supporting the use of multimodal features (gaze + head pose + speaker identity). \\
\hline
Repetitive or idiosyncratic movements & May be incorrectly interpreted as disengagement in quantitative metrics, explaining false negatives in engagement detection. \\
\hline
Intentional silence & Silence combined with gaze or subtle expressions can represent deliberate communication, highlighting the need for context-sensitive cues in modeling engagement. \\
\hline
Implications for modeling & Supports the use of multimodal features, domain adaptation, and context-aware labeling; explains why fine-tuning on IDD-specific datasets like MIDD improves performance but limitations remain. \\
\hline
\end{tabular}
\label{tab:therapist_mapping}
\end{table*}

\section{Expert Insights}
\textbf{Focus group.} To contextualize our quantitative findings, we conducted a 90-minute focus group with six therapists experienced in working with adults with IDD. The session was structured around four guiding open points that explored the naturalness of group conversation, the role of the educator, communication modalities, and the meaning of repetitive or non-verbal behaviors. Their responses highlighted the unique challenges and adaptations required for successful group interactions within this population.

\textbf{Key insights.} Therapists noted that structured group conversations are a relatively recent and evolving practice in IDD care centers. Managing turn-taking and maintaining engagement is particularly complex due to the wide heterogeneity of cognitive and behavioral profiles. While some participants dominate the floor with repeated speech acts, others remain passive or speak off-topic, resulting in unbalanced interactions. Educators often act as social scaffolds, using non-verbal cues such as gaze direction and posture to guide turn-taking. Their presence was reported as crucial, especially in early sessions when participants were unfamiliar with the task structure or the presence of observers. Without such scaffolding, several participants appeared anxious, withdrawn, or unsure of when to contribute.

Communication in these sessions extended far beyond spoken language. Many participants relied on multimodal strategies including facial expressions, physical gestures, and the use of symbolic tools like picture cards. Physical proximity and touch were also observed as legitimate and sometimes preferred channels for interaction initiation. Notably, therapists emphasized that repetitive or idiosyncratic movements (“stimming”) were not indicators of disengagement but should instead be interpreted as valid self-regulatory behaviors or expressive acts. In some cases, silence paired with gaze or subtle expression was seen as an intentional form of communication.

These observations highlight the importance of designing socially aware annotation schemes and models capable of interpreting complex, multimodal communicative acts. Visual cues alone may not suffice; successful computational modeling in IDD contexts demands the integration of gaze, body pose, prosody, and even tactile or affective cues.
Therapist insights help contextualize our quantitative findings (as shown in Table \ref{tab:therapist_mapping}), for example, by explaining the high variability in speaking activity and turn-taking as a reflection of heterogeneous participant behaviors and the guiding role of educators. 
These observations highlight that inclusive, context-aware models trained on neurodiverse datasets like MIDD can improve detection while accounting for both engagement and self-regulatory behaviors.

\begin{framed}
\noindent \textbf{Key Takeaway:} In summary, therapist observations underscore that heterogeneous behaviors, multimodal communication, and educator scaffolding shape group interactions in IDD settings, explaining patterns in speaking activity and gaze metrics, and highlighting the need for context-aware, multimodal models like those trained on MIDD to capture engagement accurately.
\end{framed}



\section{Summary and Implications}

The MIDD dataset revealed substantial differences in gaze, speaking activity, and turn-taking dynamics between neurodiverse and neurotypical participants, highlighting the limitations of models trained solely on neurotypical data. Classical and state-of-the-art eye-contact models (SVC, XGBoost, FSFNet) showed degraded performance on MIDD, though fine-tuning and multimodal feature integration improved detection accuracy. Therapist insights further contextualized these findings, emphasizing the importance of scaffolding, multimodal communication, and context-aware modeling for capturing engagement behaviors in IDD populations.

Based on these findings, we propose four data-driven strategies to enhance the development of inclusive AI systems for neurodiverse populations. 
First, \textbf{multimodal feature fusion}, e.g., integrating gaze, head pose, speaker identity, and other behavioral signals, can provide a more holistic representation of interaction. Studies in emotion recognition \cite{abdullah2021multimodal} have demonstrated that combining multiple modalities leads to more accurate and resilient models, as each modality offers complementary information about the speaker's emotional state \cite{liu2021comparing}.
Second, \textbf{domain-adaptive modeling} approaches are essential to bridge the gap between neurotypical and neurodiverse populations. Recent research \cite{tomalin2021practical, singhal2023domain} has shown that domain adaptation techniques can improve classifier performance by aligning data distributions between source and target domains, thereby enhancing generalization to neurodiverse datasets.
Finally, incorporating \textbf{context-aware and socially informed modeling}, guided by expert knowledge, can help interpret subtle communicative acts and self-regulatory behaviors. Context-aware AI systems \cite{messmer2024context}, which understand the "why" behind the data, enable more accurate interpretation of user intentions and behaviors. Furthermore, human-in-the-loop frameworks that blend AI with expert feedback \cite{natarajan2025human} can support individuals with neurodevelopmental conditions in managing digital tasks, highlighting the importance of integrating expert insights into AI system design.

\section{Conclusion}
\noindent
This work advances inclusive AI by introducing the MIDD dataset, analyzing interaction patterns in both neurodiverse and neurotypical populations, evaluating eye-contact models across these groups, and contextualizing quantitative findings with insights from experts.
Our findings highlight the importance of context-aware modeling to capture turn-taking in IDD populations, and the value of expert insights in interpreting quantitative results. Together, these contributions provide a foundation for developing AI-driven tools that support equitable, human-centered group interactions and foster social participation for neurodiverse individuals.

\addtolength{\textheight}{-3cm}   

{\small
\bibliographystyle{ieee}
\bibliography{egbib}

@article{spitale2025vita,
  title={VITA: A Multi-Modal LLM-Based System for Longitudinal, Autonomous and Adaptive Robotic Mental Well-Being Coaching},
  author={Spitale, Micol and Axelsson, Minja and Gunes, Hatice},
  journal={ACM Transactions on Human-Robot Interaction},
  volume={14},
  number={2},
  pages={1--28},
  year={2025},
  publisher={ACM New York, NY}
}

@article{tomalin2021practical,
  title={The practical ethics of bias reduction in machine translation: Why domain adaptation is better than data debiasing},
  author={Tomalin, Marcus and Byrne, Bill and Concannon, Shauna and Saunders, Danielle and Ullmann, Stefanie},
  journal={Ethics and Information Technology},
  volume={23},
  number={3},
  pages={419--433},
  year={2021},
  publisher={Springer}
}

@inproceedings{natarajan2025human,
  title={Human-in-the-loop or AI-in-the-loop? Automate or Collaborate?},
  author={Natarajan, Sriraam and Mathur, Saurabh and Sidheekh, Sahil and Stammer, Wolfgang and Kersting, Kristian},
  booktitle={Proceedings of the AAAI Conference on Artificial Intelligence},
  volume={39},
  number={27},
  pages={28594--28600},
  year={2025}
}

@inproceedings{messmer2024context,
  title={Context-aware machine learning: a survey},
  author={Messmer, Liane-Marina and Reich, Christoph and Abdeslam, Djaffar Ould},
  booktitle={Proceedings of the Future Technologies Conference},
  pages={252--272},
  year={2024},
  organization={Springer}
}

@article{singhal2023domain,
  title={Domain adaptation: challenges, methods, datasets, and applications},
  author={Singhal, Peeyush and Walambe, Rahee and Ramanna, Sheela and Kotecha, Ketan},
  journal={IEEE access},
  volume={11},
  pages={6973--7020},
  year={2023},
  publisher={IEEE}
}

@article{liu2021comparing,
  title={Comparing recognition performance and robustness of multimodal deep learning models for multimodal emotion recognition},
  author={Liu, Wei and Qiu, Jie-Lin and Zheng, Wei-Long and Lu, Bao-Liang},
  journal={IEEE Transactions on Cognitive and Developmental Systems},
  volume={14},
  number={2},
  pages={715--729},
  year={2021},
  publisher={IEEE}
}

@article{abdullah2021multimodal,
  title={Multimodal emotion recognition using deep learning},
  author={Abdullah, Sharmeen M Saleem Abdullah and Ameen, Siddeeq Y Ameen and Sadeeq, Mohammed AM and Zeebaree, Subhi},
  journal={Journal of Applied Science and Technology Trends},
  volume={2},
  number={01},
  pages={73--79},
  year={2021}
}

@article{dorfman1979formula,
  title={A formula for the Gini coefficient},
  author={Dorfman, Robert},
  journal={The review of economics and statistics},
  pages={146--149},
  year={1979},
  publisher={JSTOR}
}

@article{bigby2014conceptualizing,
  title={Conceptualizing inclusive research with people with intellectual disability},
  author={Bigby, Christine and Frawley, Patsie and Ramcharan, Paul},
  journal={Journal of applied research in intellectual disabilities},
  volume={27},
  number={1},
  pages={3--12},
  year={2014},
  publisher={Wiley Online Library}
}

@article{caforio2021design,
  title={Design issues in human-centered AI for marginalized people},
  author={Caforio, Alessandro and Pollini, Alessandro and Filograna, Antonio Salvatore and Passani, Antonella},
  year={2021}
}

@article{sigafoos2016augmentative,
  title={Augmentative and Alternative Communication (AAC) in intellectual and developmental disabilities},
  author={Sigafoos, Jeff and van der Meer, Larah and Schlosser, Ralf W and Lancioni, Giulio E and O’Reilly, Mark F and Green, Vanessa A},
  journal={Computer-assisted and web-based innovations in psychology, special education, and health},
  pages={255--285},
  year={2016},
  publisher={Elsevier}
}

@inproceedings{muller2018robust,
  title={Robust eye contact detection in natural multi-person interactions using gaze and speaking behaviour},
  author={M{\"u}ller, Philipp and Huang, Michael Xuelin and Zhang, Xucong and Bulling, Andreas},
  booktitle={Proceedings of the 2018 ACM Symposium on Eye Tracking Research \& Applications},
  pages={1--10},
  year={2018}
}

@article{regier2013dsm,
  title={The DSM-5: Classification and criteria changes},
  author={Regier, Darrel A and Kuhl, Emily A and Kupfer, David J},
  journal={World psychiatry},
  volume={12},
  number={2},
  pages={92--98},
  year={2013},
  publisher={Wiley Online Library}
}

@article{whittaker2019disability,
  title={Disability, bias, and AI},
  author={Whittaker, Meredith and Alper, Meryl and Bennett, Cynthia L and Hendren, Sara and Kaziunas, Liz and Mills, Mara and Morris, Meredith Ringel and Rankin, Joy and Rogers, Emily and Salas, Marcel and others},
  journal={AI Now Institute},
  volume={8},
  number={11},
  year={2019}
}

@article{catania2023conversational,
  title={Conversational agents in therapeutic interventions for neurodevelopmental disorders: a survey},
  author={Catania, Fabio and Spitale, Micol and Garzotto, Franca},
  journal={ACM Computing Surveys},
  volume={55},
  number={10},
  pages={1--34},
  year={2023},
  publisher={ACM New York, NY}
}

@article{jegou2018computational,
  title={A computational model for the emergence of turn-taking behaviors in user-agent interactions},
  author={J{\'e}gou, Mathieu and Chevaillier, Pierre},
  journal={Journal on Multimodal User Interfaces},
  volume={12},
  number={3},
  pages={199--223},
  year={2018},
  publisher={Springer}
}

@inproceedings{gillet2022learning,
  title={Learning gaze behaviors for balancing participation in group human-robot interactions},
  author={Gillet, Sarah and Parreira, Maria Teresa and V{\'a}zquez, Marynel and Leite, Iolanda},
  booktitle={2022 17th ACM/IEEE International Conference on Human-Robot Interaction (HRI)},
  pages={265--274},
  year={2022},
  organization={IEEE}
}

@article{do2022should,
  title={How should the agent communicate to the group? Communication strategies of a conversational agent in group chat discussions},
  author={Do, Hyo Jin and Kong, Ha-Kyung and Lee, Jaewook and Bailey, Brian P},
  journal={Proceedings of the ACM on Human-Computer Interaction},
  volume={6},
  number={CSCW2},
  pages={1--23},
  year={2022},
  publisher={ACM New York, NY, USA}
}

@article{li2023systematic,
  title={Systematic review and meta-analysis of AI-based conversational agents for promoting mental health and well-being},
  author={Li, Han and Zhang, Renwen and Lee, Yi-Chieh and Kraut, Robert E and Mohr, David C},
  journal={NPJ Digital Medicine},
  volume={6},
  number={1},
  pages={236},
  year={2023},
  publisher={Nature Publishing Group UK London}
}

@inproceedings{mueller2018lowrapport,
    author = {M\"{u}ller, Philipp and Huang, Michael Xuelin and Bulling, Andreas},
    title = {Detecting Low Rapport During Natural Interactions in Small Groups from Non-Verbal Behaviour},
    year = {2018},
    isbn = {9781450349451},
    publisher = {Association for Computing Machinery},
    address = {New York, NY, USA},
    url = {https://doi.org/10.1145/3172944.3172969},
    doi = {10.1145/3172944.3172969},
    booktitle = {Proceedings of the 23rd International Conference on Intelligent User Interfaces},
    pages = {153–164},
    numpages = {12},
    location = {Tokyo, Japan},
    series = {IUI '18}
}

@inproceedings{mueller2018eyecontact,
    author = {M\"{u}ller, Philipp and Huang, Michael Xuelin and Zhang, Xucong and Bulling, Andreas},
    title = {Robust eye contact detection in natural multi-person interactions using gaze and speaking behaviour},
    year = {2018},
    isbn = {9781450357067},
    publisher = {Association for Computing Machinery},
    address = {New York, NY, USA},
    url = {https://doi.org/10.1145/3204493.3204549},
    doi = {10.1145/3204493.3204549},
    booktitle = {Proceedings of the 2018 ACM Symposium on Eye Tracking Research \& Applications},
    articleno = {31},
    numpages = {10},
    location = {Warsaw, Poland},
    series = {ETRA '18}
}

@inproceedings{ma2024less,
    author = {Ma, Fuyan and He, Yiran and Sun, Bin and Li, Shutao},
    title = {Less is More: Adaptive Feature Selection and Fusion for Eye Contact Detection},
    year = {2024},
    isbn = {9798400706868},
    publisher = {Association for Computing Machinery},
    address = {New York, NY, USA},
    url = {https://doi.org/10.1145/3664647.3688987},
    doi = {10.1145/3664647.3688987},
    booktitle = {Proceedings of the 32nd ACM International Conference on Multimedia},
    pages = {11390–11396},
    numpages = {7},
    location = {Melbourne VIC, Australia},
    series = {MM '24}
}

@inproceedings{openface,
    author={Baltrušaitis, Tadas and Robinson, Peter and Morency, Louis-Philippe},
    booktitle={2016 IEEE Winter Conference on Applications of Computer Vision (WACV)}, 
    title={OpenFace: An open source facial behavior analysis toolkit}, 
    year={2016},
    volume={},
    number={},
    pages={1-10},
    doi={10.1109/WACV.2016.7477553}
}

@article{Antaki2012idadults,
    author = {Antaki, Charles},
    year = {2012},
    month = {05},
    pages = {},
    title = {Two conversational practices for encouraging adults with intellectual disabilities to reflect on their activities},
    volume = {57},
    journal = {Journal of intellectual disability research : JIDR},
    doi = {10.1111/j.1365-2788.2012.01572.x}
}

@article{Wallin1797526,
    author = {Wallin, Sofia and Hemmingsson, Helena and Thunberg, Gunilla and Wilder, Jenny},
    doi = {10.1080/07434618.2023.2243517},
    institution = {Stockholm University, Department of Special Education},
    journal = {Augmentative and Alternative Communication: AAC},
    number = {1},
    pages = {19--30},
    title = {Turn-taking and communication modes of students and staff in group activities at non-inclusive schools for students with intellectual disability},
    volume = {40},
    year = {2024}
}

@article{Tepencelik2024autism,
    author = {Tepencelik, Onur Necip and Wei, Wenchuan and Luo, Mirabel and Cosman, Pamela and Dey, Sujit},
    day = {12},
    doi = {10.2196/55339},
    issn = {2561-326X},
    journal = {JMIR Form Res},
    month = {Aug},
    pages = {e55339},
    title = {Behavioral Intervention for Adults With Autism on Distribution of Attention in Triadic Conversations: A/B-Tested Pre-Post Study},
    url = {https://doi.org/10.2196/55339},
    volume = {8},
    year = {2024}
}

@article{antaki2017profound,
    author    = {Antaki, Charles and Crompton, Rebecca J. and Walton, Chris and Finlay, W. M. L.},
    title     = {How adults with a profound intellectual disability engage others in interaction},
    journal   = {Sociology of Health \& Illness},
    volume    = {39},
    number    = {4},
    pages     = {581--598},
    year      = {2017},
    doi       = {10.1111/1467-9566.12500},
    publisher = {Wiley}
}

@misc{mazzola2023icub,
    author       = {Mazzola, Carlo and Rea, Francesco and Sciutti, Alessandra},
    title        = {Real‐Time Addressee Estimation: Deployment of a Deep‐Learning Model on the iCub Robot},
    note         = {Unpublished},
}
}

\end{document}